\documentclass[12pt]{article}
\usepackage{amsmath}
\usepackage{graphicx,psfrag,epsf}
\usepackage{enumerate}
\usepackage{algorithm}
\usepackage{algorithmic}
\usepackage{float} 
\usepackage{url} 
\usepackage{cite}
\usepackage{array} 
\usepackage{amssymb}

\newcommand{\blind}{0}

\begin{document}

\def\spacingset#1{\renewcommand{\baselinestretch}%
{#1}\small\normalsize} \spacingset{1}


\if0\blind
{
  \title{\bf Effective Calibration Transfer via M\"{o}bius and Affine Transformations}
  \author{Casey Kneale, Karl S. Booksh\hspace{.2cm}\\
    Department of Chemistry and Biochemistry\\
    University of Delaware Newark Delaware 19716\\
	}
  \maketitle
} \fi

\if1\blind
{
  \bigskip
  \bigskip
  \bigskip
  \begin{center}
    {\LARGE\bf Effective Calibration Transfer via Mobius and Affine Transformations}
\end{center}
  \medskip
} \fi

\bigskip
\begin{abstract}
A novel technique for calibration transfer called the Modified Four Point Interpolant (MFPI) method is introduced for near infrared spectra. The method is founded on physical intuition and utilizes a series of quasiconformal maps in the frequency domain to transfer spectra from a slave instrument to a master instrument\textsc{\char13}s approximated space. Comparisons between direct standardization (DS), piecewise direct standardization (PDS), and MFPI for two publicly available datasets are detailed herein. The results suggest that MFPI can outperform DS and PDS with respect to root mean squared errors of transfer and prediction. Combinations of MFPI with DS/PDS are also shown to reduce predictive errors after transfer.
\end{abstract}

\noindent%
{\it Keywords:} Calibration Transfer; Near Infrared (NIR); Four Point Interpolant (FPI); Direct Standardization (DS); Piecewise Direct Standardization (PDS).

\newpage
\spacingset{1.45} 
\section{Introduction}
\label{sec:intro}
Many calibration models rely on trends of small variations in response for efficacy. Although these responses are correlated to chemical features, they are not pure representations of the feature itself. When a feature of an analyte is detected from different instruments the shape, magnitude, and location of the responses are not identical. In most circumstances, two different instruments cannot use the same model and obtain commensurate results. This problem is known as calibration transfer.

The goal of calibration transfer is to transfer data obtained from a secondary or slave instrument to a space defined by a primary or master instrument. This allows for the same calibration model to be applied, ideally, without change or loss of quality to any secondary instrument. The difficulty lies within accounting for each instrument\textsc{\char13}s unique contributions to random and systematic error.

It is believed that the instrumental deviations for near infrared (NIR) spectroscopy, the primary focus of this work, originate from artifacts incurred by stray light. Stray light is thought to introduce a multiplicative effect which causes linear deviations in baselines. The method of direct standardization (DS) was developed to correct these kinds of linear responses. DS constructs a full rank standardization matrix by multiplication of the inverse slave calibration spectra with the master spectra in the PCA column space defined by the master\textsc{\char13}s set  \cite{Wang1991}. This matrix allows for any similar spectra collected by a slave instrument to be transformed into the master instrument\textsc{\char13}s approximate space as follows: PCA transformation to masters space, multiplication with standardization matrix, and deprojecting back into instrument space.
 
It was found that the construction of the standardization matrix by concatenation of piecewise spectral DS matrices often afforded a better estimate of the masters\textsc{\char13} space. This method is referred to as PDS. It is widely considered the most universal transformation technique because it accounts for the inhomogeniety of the deviations across wavelength regions. Wang et. al, displayed compelling empirical results which showed that the effect of stray light is structured. The results suggested that calibration transfer models can benefit from mean centering and an additive mean correction term\cite{Wang1995}. Such a modification is applicable to both DS and PDS methods.

The most widely used transformation techniques to transfer calibration models are DS and PDS despite many other efforts and suggestions. Although these methods are physically grounded and mathematically well reasoned, they still lack the ability to overcome instrumental deviations which are more ill-posed. Some measured artifacts which are non-linear with respect to wavelength vs absorption could originate from the following: optical losses incurred by path elements, alignment errors, and different detection schemes. The merit of the work depicted herein is an alternative transformation approach to calibration transfer which attempts to account for these discrepancies and is based on physical reasoning.

\section{Theory}
\label{sec:desc}
The frequency domain provides fertile ground for transferring instrumental response. High frequency noise and low frequency backgrounds are well resolved from the perspective of interferograms. A frequency coefficient can be seen as a point in 3 space; the frequency itself is real and the coefficient lays in the Argand plane. In the case of Fourier Transform NIR instruments, each bin represents a difference in optical path-length and the respective modulus is directly proportional to the convolution of detected photon amplitudes or energy. For dispersive instruments, such a statement is not true but a similar more abstract analogy holds. 

Progress in elementary optics has shown that transfer or scattering matrices can be developed that accurately model the effects of various materials onto a given field or amplitude spectrum of light. These effects may be changes in optical phase, amplitude, or polarization. M\"{o}bius transformations, also referred to as Linear Fractional Transforms (LFT), allow the construction of these matrices and their manipulations to be convenient and stable\cite{Sirenko}. These matrices allow for the interconversion of responses from one optical path to another. 

Unfortunately, the differences between two instruments cannot be readily regressed to a single scattering matrix. Such an effort would require commercial NIR instruments to feature variable incident angles, polarization control, known calibration parameters of detectors, etc. These interrogations are required because currently the only means to measure light is through spatial considerations of intensity measures. Overall a loss of generality and practicality is inevitable by pursuing a completely physical solution to the calibration transfer problem. 

Instead, an interpolated transformation scheme which abides by the geometry of the aforementioned physical reasoning may be applied. Transforms and interpolants should conform to the cylindrical nature of interoferogram space. In the frequency domain, instrumental deviations may be represented by subtends in the complex plane, and their associated magnitudes at discretized bins.

One such transformation technique was introduced by Lipman, et al. for purposes of user-friendly computer graphic manipulations\cite{lipman}. In its original form, the Four Point Interpolant method, begins by creating two LFTs which maps two sets of four points from their respective problem spaces to parallelograms. Then an optimized bijective affine map is constructed which bridges the two parallelogram spaces while minimizing maximal conformal distortion (Figure 1). Their method has been shown to perform circular interpolations which do not cross themselves, preserve angles between points throughout the transformation(s), and bolster mathematical rigor to support its efficacy. Pseudo code and mathematical proofs for FPI are provided in the original publication. However, the authors would like to mention a potential nuance in their published work; the authors state to solve a quadratic equation using the 3$^{rd}$ and 4$^{th}$ singular values. In most programming languages (R, Octave, etc) the singular values they refer to are organized as the 5$^{th}$ and 6$^{th}$.

\begin{figure}[H]
\begin{center}
\includegraphics[width=3in]{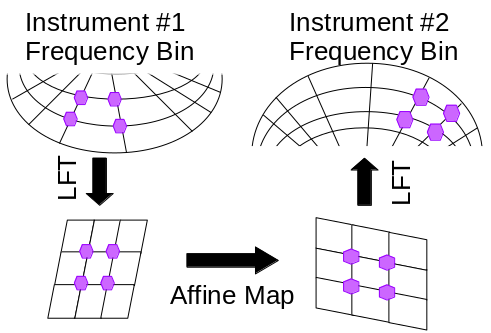}
\end{center}
\caption{Flow diagram of the FPI. \label{fig:first}}
\end{figure}

When applied to spectra, a set of four sample points in a frequency bin can define a parallelogram space for a given instrument. To improve the circularity of the interoferogram space, the Cayley transform should be implemented.The Cayley transform need not preserve its sign to return effective results as is done in traditional optics on scattering matrices. This space and all points along a respective transformed frequency bin are then interpolated to a parallelogram, affine mapped to the master instruments parallelogram space, inverse transformed to the master\textsc{\char13}s Cayley space, and finally inverse Cayley transformed to the masters approximated interoferogram space. This procedure details the modified four point interpolant (MFPI) method.

The generation of the interpolant maps for the MFPI algorithm is based on minimizing an objective function. In this work, the four observations of the most similar interoferograms between instruments are selected at each bin. The points are cycled in ascending order with respect to their differences such that the lowest root mean squared error of the transformed interoferograms is attained (Supplementary Algorithm 1). This preference allows for mappings of similar interoferograms to be constructed.  

\section{Data}
\label{sec:data}
In all cases presented, no samples were considered outliers, the entire spectral range was utilized, pre-processing and post-processing were not performed. 

For convenience and reproducibility, the International Diffuse Reflectance Conference Shoot-out 2002 pharmaceutical data set \cite{IDRC} was employed as a primary investigation of the MFPI method for calibration transfer. The data set features 155 calibration, 40 validation, and 460 test samples which were analyzed by two different NIR spectrometers. The data was collected with 2 nm resolution. Three property values for model construction are also contained. 

A secondary dataset, the Cargill-Corn data \cite{Corn}, was also investigated. The Cargill-Corn data features 3 NIR instruments (M5, MP5, and MP6) of the same resolution (2 nm) and 80 spectra each. The Corn data features four property values.

\section{Results}
\label{sec:Res}
DS, PDS, and MFPI were all applied to the validation and test set of the pharmaceutical data. The MFPI algorithm rejected one sample from the set as it did not contribute to a lower error. The PDS window size was empirically set to 300 bins. The window size was found by incrementing from 3-310 bins and keeping the size which resulted in the lowest error. The subtraction of the transformed slave spectra from their respective sample body offered a visual means to assess the efficacy of the algorithms (Figure 2). 

It was perceived that DS had the worst performance of the three methods employed. While PDS and DS appeared commensurate for the validation and test sets. A figure of merit, mean root mean square error of transfer (MRMSET), was utilized to gain empirical insight into the similarity of the transformed slave spectra and that of the unaltered master spectra.  
\begin{figure}[t]
	\begin{center}
		\includegraphics[width=4in]{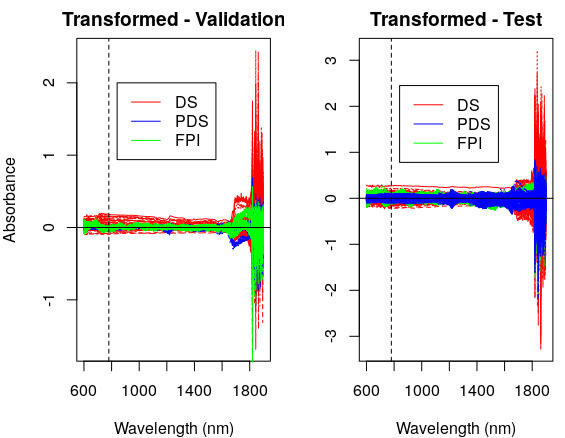}
	\end{center}
	\caption{Difference spectra of the transferred data minus the master's validation data (left) and the transformed data minus the test data (right). \label{fig:second}}
\end{figure}

We define the mean root mean square error of transfer as, 

$MRMSET =  \frac{1}{M} \sum_{\lambda = 0} ^{M}  \sqrt{\frac{1}{N} \sum_{n=0}^{N} \Re(M_{\lambda, n} - T_{\lambda, n})^2 + \Im(M_{\lambda, n} - T_{\lambda, n})^2}$. Where M denotes the master instruments spectra, T denotes the transformed spectra, N is the number of samples, and M is the number of wavelengths.
 
The MRMSET figure of merit demonstrated that the MFPI and PDS algorithms offered calibration transfers which were most similar to their master spectra (Table I). It was found that MFPI had the lowest MRMSET across the entire spectral region for the validation set. PDS was 9.7\% and 3.6\% different from MFPI for the transfer of instrument 1 to 2 and 2 to 1, respectively. For the test set however, PDS (1.15, 1.16) featured lower MRMSETs than MFPI (1.19, 1.26). When MFPI used every other sample (N = 40) in the calibration set to build it’s interpolating schema, the MRMSET for validation and test sets were reduced (Supplementary Table 1). Methodological differences were assessed via Bland-Altman analysis.

\begin{table}[htb]
	\begin{center}
		\caption{Root mean squared errors of transfer for the calibration (MRMSECT), validation (MRMSEVT), and test (MRMSETT) sets. Calibration transfers from instrument 2 to 1 and 1 to 2 are denoted with a and b respectively.} \label{tab:tabtwo}
		\begin{tabular}{r|ccc}
			& MRMSECT & MRMSEVT & MRMSETT   \\\hline
			\textbf{DS$^a$}  & $1.051 \cdot 10^{-13}$ & $0.188$ & $1.980$ \\
			\textbf{PDS$^a$}  & $0.407$ & $0.097$ & $1.145$ \\
			\textbf{FPI$^a$}  & $0.407$ & $0.088$ & $1.193$ 
			\\
			\textbf{DS$^b$}  & $1.598 \cdot 10^{-13}$ & $0.162$ & $1.858$ \\
			\textbf{PDS$^b$}  & $0.399$ & $0.084 $ & $1.157$ \\
			\textbf{MFPI$^b$}  & $0.417$ & $0.081$ & $1.261$ \\
		\end{tabular}
	\end{center}
\end{table}

Bland-Altman analysis of the average transformed spectra vs the average master spectra revealed differences in bias and precision between the techniques (Figure 3). In all cases the method with the least bias was DS. For both the validation and test sets PDS afforded positive biases while MFPIs varied by transfer. In order of increasing span of their limits of agreement or decreasing precision the methods were ranked as follows: PDS, DS, MFPI. The lack of precision in the MFPI method was hypothesized to be a direct result of noise. High frequency features were commonly observed in the MFPI transformed spectra (Supplementary Figure 1). This was believed to be an artifact of interpolating in the frequency domain. 
\begin{figure}[H]
	\begin{center}
		\includegraphics[width=4.0in]{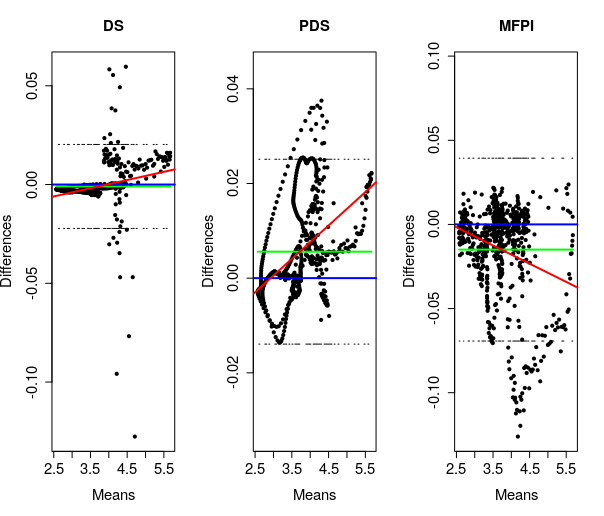}
	\end{center}
	\caption{Bland-Altman plots of the pharmaceutical data's test set after DS (left), PDS (center), and MFPI (right). \label{fig:third}}
\end{figure}

The transferred spectra were utilized to determine if the noise from MFPI effected the performance of principal component regression (PCR) models on the three available property values (Table II). Leave-one-out cross-validation was used for reproducability. The transfers of spectra from instrument 2 to instrument 1 and instrument 1 to instrument 2 were both considered. The number of principal components employed for the calibration models were the number where the minima in root mean squared error of prediction (RMSEP) was observed.

We define the root mean square error of prediction as, 

$RMSEP =  \sqrt{\frac{1}{N} \sum_{n=0}^{N} (Y_{ n} - \hat{Y}_{n})^2}$. Where $Y_{ n}$ denotes the reference property value, $\hat{Y}_{n}$ denotes the predicted property value, and N is the total number of observations.

\begin{table}[H]
	\begin{center}
		\caption{Root mean squared errors of prediction for the pharmaceutical validation, and test sets. Calibration transfers from instrument 2 to 1 and 1 to 2 are denoted with a and b respectively. Methods which afforded the lowest RMSEP for a respective
		transfer from master $\rightarrow$ slave are emphasized with bold text.} \label{tab:tabthree}
		\begin{tabular}{c|cc|cc|cc}
			& RMSEVP & RMSETP  & RMSEVP  & RMSETP & RMSEVP & RMSETP \\
			& (weight) & (weight) & (hardness) & (hardness) & (assay) & (assay)  \\\hline  
			\textbf{Master$^a$}  & $3.730$ & $5.353$ & $1.202$ & $0.969$& $3.978$& $4.715$ \\
			\textbf{Inst 2.$^a$} & $3.591$ & $5.801$ & $0.871$ & $1.123$& $5.329$& $7.995$ \\
			\textbf{DS$^a$}  & \textbf{3.303} & $5.677$ & $1.165$ & $1.062$& \textbf{3.236} & \textbf{6.027} \\
			\textbf{PDS$^a$}  & $4.126$ & \textbf{4.994} & $1.166$ & \textbf{0.948}& $7.743$& $15.08$ \\
			\textbf{MFPI$^a$}  & $3.687$ & $5.534$ & \textbf{1.155} & $0.954$& $4.506$& $8.654$ \\ \hline

			\textbf{Master$^b$}  & $3.787$ & $5.776$ & $1.256$ & $0.978$& $3.620$& $4.919$ \\
			\textbf{Inst 1.$^b$} & $5.660$ & $5.219$ & $1.788$ & $1.172$& $13.87$& $12.54$ \\
			\textbf{DS$^b$}  & $6.948$ & $5.045$ & $1.507$ & $1.079$& $5.318$& \textbf{5.329} \\
			\textbf{PDS$^b$}  & \textbf{2.853}  & \textbf{3.831} & \textbf{1.108} & $0.988$& $7.895$& $14.97$ \\
			\textbf{MFPI$^b$}  & $3.069$ & $4.186$ & $1.293$ & \textbf{0.934}& \textbf{2.703}& $6.183$ 
		\end{tabular}
	\end{center}
\end{table}    

Stronger and more deleterious contributions from interpolation where observed in PCR models constructed from the Cargill-Corn data. Crudely optimized, manual frequency bin selection was applied to all of the MFPI transfers in a uniform manner to dampen these artifacts. Combinations of the MFPI, PDS, and DS techniques were also assessed. The Kennard-Stone algorithm was utilized on half of the available samples (N=40) to create a calibration set. The remaining samples were all considered a test set. The RMSEPs obtained for predicting on the uncalibrated samples for every instrument combination are provided below (Table III) and in the supplementary information.

\begin{table}[H]
	\begin{center}
		\caption{Average root mean squared errors of prediction for PCR models developed on the Cargill-Corn's Starch property. Methods which afforded the lowest RMSEP for a respective transfer from master $\rightarrow$ slave are emphasized with bold text.} \label{tab:tabfour}
		\resizebox{\textwidth}{!}{
		\begin{tabular}{lc|c|c|c|c|c}
			&M5 $\rightarrow$ MP5 &MP6 $\rightarrow$ MP5 &MP5 $\rightarrow$ MP6 &M5 $\rightarrow$ MP6 &MP6 $\rightarrow$ M5 &MP5 $\rightarrow$ M5
			 \\\hline  
\textbf{Master}	&1.449	&1.449	&0.171	&0.171	&3.515	&3.515\\
\textbf{Slave}	&3.898	&9.343	&2.668	&3.129	&10.192	&1.947\\
\textbf{DS}	&1.498	&1.452	&0.625	&\textbf{0.495}	&3.816	&3.404\\
\textbf{PDS}	&1.527	&2.035	&0.804	&0.854	&4.038	&3.426\\
\textbf{MFPI}	&2.163	&4.607	&1.671	&2.665	&\textbf{0.955}	&\textbf{0.889}\\
\textbf{MFPI-DS}	&\textbf{1.392}	&\textbf{1.328}	&\textbf{0.513}	&0.752	&3.692	&3.447\\
\textbf{MFPI-PDS}	&1.916	&4.625	&1.841	&1.459	&3.943	&4.142\\
			
		\end{tabular}
	}
	\end{center}

\end{table}

\section{Discussion}
\label{sec:Discussion}

The MRMSET figures of merit obtained for the pharmaceutical data provided evidence of overfitting when all 80 calibration samples were made available to the MFPI algorithm. This test demonstrated that the MFPI algorithm was amenable to subsampling; this is an important feature for any calibration transfer algorithm. In practice it is highly preferable to utilize fewer standard samples on slave instruments to facilitate transfers of the same or higher quality. This attribute is known as subsampling and has been well established for the DS and PDS methods. 

However, the RMSEPs for most of the PCR models increased when only every other sample was utilized (Supplementary Table I). This result suggested that the MFPI algorithm was overfit by complete access to the calibration set, but the improvements in spectral similarities were not necessarily occurring at the regions of interest for the PCR models.

For the pharmaceutical data, MFPI can be seen as a complimentary calibration transfer technique. MFPI, DS, and PDS each had several transfers where they built a model with the lowest RMSEP. Of the transfer methods, MFPI always possessed either the lowest or second lowest RMSEP. This observation held true even when every other sample in the calibration data were utilized to build the transfer subset (Supplementary Table II). This observation provided evidence which suggested that the perceived noise from the interpolation in the frequency domain did not substantially hinder the efficacy of the technique. However, post-processing could very well improve these results.

Each of the three methods had several instances where their PCR RMSEPs were greater then the raw spectra of the slave instrument. Such an event was considered by the authors to be a failed transfer. PDS exhibited 5 instances of such transfers; DS possessed 2, and MFPI had 3. 

From this study, it would be unfair to consider a method which featured a larger predictive error than its respective slave spectra ineffective as a whole. The lack of preprocessing, simple leave-one-out cross-validation, and facile models utilized herein were intentionally designed as to not favor any of the calibration transfer techniques. However, RMSEPs where the slave instrument spectra performed better then transferred spectra are contradictory to the goals of calibration transfer. 

The methods were also tested on the Cargill-Corn data because it had different spectral features and fewer samples than the pharmaceutical data. When all of the frequency bins were interpolated by MFPI, the calibration models suffered largely from high frequency artifacts. Frequency bin selection was thus performed to attempt the removal of frequency ranges which inhibited model development while still allowing interpolation of bins which were beneficial. The optimal means to approach the problem of bin selection is an area of active research.

The selection of samples to construct the calibration sets resulted in a potential conflict of scientific honesty. Sample subsets which were optimized for the MFPI method often resulted in sub-par transfers for either of the direct standardization methods. Similarly, the employment of individual sample subsets for each method would introduce a more open interpretation of results due to sample selection efficacy. The pursuit of fair treatment for all of the methods was facilitated via the uniform employment of a euclidean distance Kennard-Stone design. It must be stated, there was no evidence which suggested that a Kennard-Stone design would be ideal for the MFPI method as it often is for DS methods.

There were 24 possible calibration models (6 instrument transfers, 4 property values) which could be built from the Cargill-Corn data. Of the possible transfer scenarios, 13 which utilized MFPI (5 - MFPI, 8 - MFPI-DS) resulted in the lowest RMSEPs. Although there was only a two sample advantage between the standard techniques and those which used MFPI, MFPI was beneficial for certain transfers. In some cases there were large differences in performance between competing methods. For example, the RMSEPs obtained from both the moisture and starch properties after the MFPI transfers (MP6 and MP5 to instrument M5) were found to be an order of magnitude lower than PDS/DS. 

In several transfers, the coupling of direct standardization techniques to MFPI resulted in the lowest RMSEP. In 3 transfers, the lowest RMSEPs obtained by coupled techniques were less than 5\% different from the native method. However, there were more cases where the coupled techniques performed the best and were greater than 5\% different. For example, all three of the values for which DS-MFPI transferred the starch property with the least error were greater than 25\% different from the DS method alone. 

The utility of MFPI alone or in conjunction with DS or PDS can be seen as beneficial. Whether or not the method can account for instrumental deviations which DS/PDS cannot requires further testing on controlled data sets. Like any other technique in chemometrics, the employment of MFPI appears to be context dependent. For example, MFPI and MFPI-DS tended to out perform the other techniques for the starch property, but for the protein property, direct standardization was more often better suited.

\section{Future Work}
\label{sec:Future}
The novelty of the MFPI approach leaves many areas of research open for future investigations. An investigation of what type of samples, or design of experiment, should be utilized in order to obtain the best calibration transfer is pertinent. The MFPI method described in this work utilized four frequency bins which were the most similar across spectra while offering the lowest RMSE of the transformed interoferograms. It is hypothesized that the incorporation of sample replicates to capture intersample variabilities would be beneficial in obtaining more precise transfers. A four point window of replicate measurand and instrumental variance between two similar samples could weight modeling power in a way that it more effectively approximates transfer.

We hope to assess the possibility of an analyte independent calibration transfer using a methodology similar to the MFPI. The overall instrumental variations in measuring common superpositions of radiation may be accounted for with a similar approach. Such an effort would allow for many models to be shared from only one calibration set. Data collection for both of these hypothesis’ have already begun in our laboratory

\section{Conclusion}
\label{sec:Conclusion}
MFPI is a new technique for calibration transfer that is based on quasiconformal mappings of interoferograms. MFPI furnished either the lowest or second lowest RMSEP of transferred spectra in the pharmaceutical data set. Like DS and PDS, the new technique is amenable to subsampling. In the Cargill-Corn data set, distinct reductions in modeling error were observed for methods which employed MFPI. The ability to select frequency bins of interest for interpolation allowed for the reduction of interpolation artifacts, effective calibration transfers, and symbiosis with classic standardization techniques.

\section{Acknowledgements}
\label{sec:Acknowledge}
The authors would like to thank Dominic Poerio for discussions about comparative test cases, and Vicki Lyn Wallace for revising the manuscript and criticism of the original algorithm.

\newpage 
\begin{center}
	{\large\bf SUPPLEMENTARY MATERIAL}
\end{center}

\begin{algorithm}[H]
	\caption{MFPI Map Construction}
	\begin{algorithmic}
		\REQUIRE $Master_{M,N}$, $Slave_{M,N}$, tolerance, reject, maxIter
		
		\ENSURE $ForwardLFT_{M,2,2}$, $BackwardLFT_{M,2,2}$, $AffineParams_{\lambda, 4}$
		
		\STATE $difference_{M,N} \leftarrow \sqrt{\Re(Master - Slave)^2 + \Im(Master - Slave)^2}$
		
		\FOR{$\lambda \in \{1,\dots,M\}$}
		\STATE \#Create 4 points from nearest samples
		\FOR{$s \in \{1,\dots,4\}$}
		\STATE $sample_s \leftarrow which.min(difference[\lambda,])$
		\STATE $difference[\lambda, sample_s] \leftarrow difference[\lambda, which.max(difference[\lambda,])]$
		\ENDFOR
		
		\STATE $iter \leftarrow 0$
		\STATE $BestRMSE \leftarrow tolerance$
		
		\WHILE{$curRMSE >= tolerance$ $\&$ $iter < maxIter$}
		\STATE $i \leftarrow which(c(1,2,3,4) == (iter\%\%4) )$
		
		\STATE $sample_i \leftarrow which.min(difference_{\lambda,1:N})$
		
		\STATE $difference_{\lambda, i} <- difference[\lambda, which.max(difference[\lambda,])$
		\STATE \#Create FPI maps
		\STATE $FPIparams \leftarrow FPI(Slave[\lambda, CCWSort(sample)], Master[\lambda, CCWSort(sample)])$

		\STATE $Transformed \leftarrow TransformByFPI( Slave[\lambda, ], FPIparams)$
		
		\STATE $CurRMSE \leftarrow RMSE(Master[\lambda,], Transformed)$
		\STATE \#Replace parameters with the ones which give lowest error
		\IF{ $CurRMSE < BestRMSE$}
		\STATE $BestRMSE \leftarrow CurRMSE$
		\STATE $ForwardLFT[\lambda,,] \leftarrow FPIparams\$ForwardLFT$
		\STATE $BackwardLFT[\lambda,,] \leftarrow FPIparams\$BackwardLFT$
		\STATE $AffineParams[\lambda, ] \leftarrow c(FPIparams\$w,  FPIparams\$z, FPIparams\$l)$
		\ENDIF
		
		\ENDWHILE
		
		\ENDFOR
		
	\end{algorithmic}
\end{algorithm}

\setcounter{figure}{0}   
\setcounter{table}{0}

\begin{table}[htb]
	\begin{center}
		\caption{Root mean squared errors of transfer for the calibration (MRMSECT), validation (MRMSEVT), and test (MRMSETT) sets. Calibration transfers from instrument 2 to 1 and 1 to 2 are denoted with a and b respectively.} \label{tab:tabtwo}
		\begin{tabular}{r|ccc}
			& MRMSECT & MRMSEVT & MRMSETT   \\\hline
\textbf{DS$^a$}	&0.292	&0.155	&1.724\\
\textbf{PDS$^a$} &0.409	&0.096	&1.145\\
\textbf{MFPI$^a$}&0.198	&0.096 &1.182	
			\\
\textbf{DS$^b$}	&0.311	&0.160	&1.821\\
\textbf{PDS$^b$}	&0.400	&0.083	&1.157\\
\textbf{MFPI$^b$}	&0.201	&0.083	&1.225	\\
		\end{tabular}
	\end{center}
\end{table}

\begin{figure}[H]
	\begin{center}
		\includegraphics[width=4in]{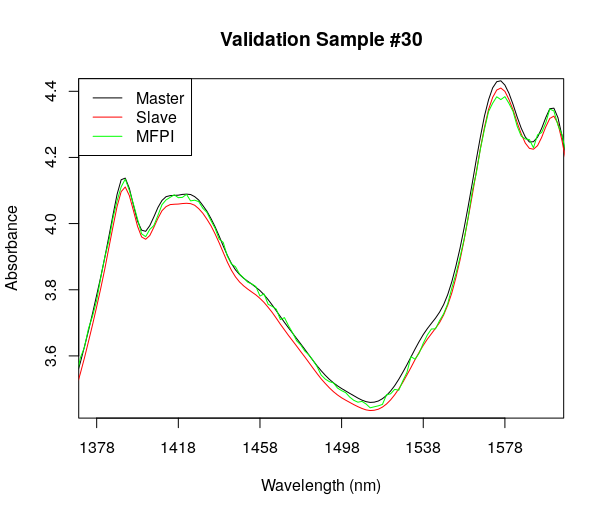}
	\end{center}
	\caption{Exemplary spectral over-lay of the MFPI method, the master spectra, and the slave. High frequency noise can be seen in the MFPI transferred data. \label{fig:second}}
\end{figure}

\begin{table}[ht]
	\begin{center}
		\caption{Root mean squared errors of prediction for every other sample in the pharmaceutical validation, and test sets. Calibration transfers from instrument 2 to 1 and 1 to 2 are denoted with a and b respectively. Methods which afforded the lowest RMSEP for a respective
			transfer from master $\rightarrow$ slave are emphasized with bold text.} \label{tab:tabthree}
		\begin{tabular}{c|cc|cc|cc}
& RMSEVP & RMSETP  & RMSEVP  & RMSETP & RMSEVP & RMSETP \\
& (weight) & (weight) & (hardness) & (hardness) & (assay) & (assay)  \\\hline  
\textbf{Master$^a$}	&3.730	&5.353	&1.202	&0.969	&3.978	&4.715\\
\textbf{Inst 2.$^a$}	&3.591	&5.810	&0.871	&1.123	&5.329	&7.995\\
\textbf{DS$^a$}	&\textbf{3.586}	&5.641	&1.074	&\textbf{1.024}	&\textbf{4.232}	&\textbf{6.023}\\
\textbf{PDS$^a$}	&4.058	&\textbf{5.102}	&1.160	&\textbf{0.950}	&6.606	&15.013\\
\textbf{MFPI$^a$} &3.806	&5.233	&1.230	&0.955	&5.041	&7.417\\ \hline

\textbf{Master$^b$}	&3.787	&5.776	&1.256	&0.978	&3.620	&4.919\\
\textbf{Inst 1.$^b$}	&5.660	&5.219	&1.788	&1.172	&13.86	&12.54\\
\textbf{DS$^b$}	&6.555	&5.607	&1.439	&1.068	&4.842	&\textbf{5.437}\\
\textbf{PDS$^b$}	&\textbf{2.975}	&\textbf{3.912}	&\textbf{1.127}	&0.992	&6.714	&14.81\\
\textbf{MFPI$^b$}	&3.988	&4.553	&1.308	&\textbf{0.976}	&\textbf{4.252}	&6.010\\

		\end{tabular}
	\end{center}
\end{table}

\begin{table}[h]
	\begin{center}
		\caption{Average root mean squared errors of prediction for the Cargill-Corn's Moisture property. } \label{tab:tabfive}
		\resizebox{\textwidth}{!}{
			\begin{tabular}{lc|c|c|c|c|c}
			&M5 $\rightarrow$ MP5 &MP6 $\rightarrow$ MP5 &MP5 $\rightarrow$ MP6 &M5 $\rightarrow$ MP6 &MP6 $\rightarrow$ M5 &MP5 $\rightarrow$ M5
				\\\hline  
\textbf{Master}	&0.177	&0.177	&0.011	&0.011	&2.091	&2.091\\
\textbf{Slave}	&0.188	&1.463	&1.491	&1.671	&4.858	&1.191\\
\textbf{DS}	&\textbf{0.204}	&\textbf{0.167}	&0.199	&\textbf{0.191}	&2.281	&2.038\\
\textbf{PDS} &0.276	&0.390	&0.412	&0.352	&1.943	&2.115\\
\textbf{MFPI} &0.293	&0.611	&0.449	&0.518	&\textbf{0.384}	&\textbf{0.570}\\
\textbf{MFPI-DS} &0.256	&0.190	&\textbf{0.198}	&0.239	&2.299	&2.122\\
\textbf{MFPI-PDS} &0.810	&0.967	&0.308	&0.271	&2.153	&2.265\\
				
			\end{tabular}
		}
	\end{center}
	
\end{table}

\begin{table}[h]
	\begin{center}
		\caption{Average root mean squared errors of prediction for the Cargill-Corn's Protein property. } \label{tab:tabsix}
		\resizebox{\textwidth}{!}{
			\begin{tabular}{lc|c|c|c|c|c}
			&M5 $\rightarrow$ MP5 &MP6 $\rightarrow$ MP5 &MP5 $\rightarrow$ MP6 &M5 $\rightarrow$ MP6 &MP6 $\rightarrow$ M5 &MP5 $\rightarrow$ M5
				\\\hline  
	\textbf{Master}	&0.356	&0.356	&0.136	&0.136	&0.390	&0.390\\
	\textbf{Slave}	&1.462	&2.077	&0.477	&0.272	&1.207	&0.271\\
	\textbf{DS}	&\textbf{0.372}	&0.399	&0.178	&\textbf{0.204}	&\textbf{0.447}	&\textbf{0.378}\\
	\textbf{PDS}	&0.506	&0.591	&0.508	&0.507	&0.618	&0.543\\
	\textbf{MFPI}	&1.126	&0.994	&0.987	&0.799	&0.624	&0.506\\
	\textbf{MFPI-DS}	&0.448	&\textbf{0.374}	&\textbf{0.169}	&0.209	&0.505	&0.416\\
	\textbf{MFPI-PDS}	&0.704	&1.006	&0.527	&0.531	&0.678	&0.957\\
				
			\end{tabular}
		}
	\end{center}
	
\end{table}
\begin{table}[h]
	\begin{center}
		\caption{Average root mean squared errors of prediction for the Cargill-Corn's Oil property. } \label{tab:tabseven}
		\resizebox{\textwidth}{!}{
			\begin{tabular}{lc|c|c|c|c|c}
			&M5 $\rightarrow$ MP5 &MP6 $\rightarrow$ MP5 &MP5 $\rightarrow$ MP6 &M5 $\rightarrow$ MP6 &MP6 $\rightarrow$ M5 &MP5 $\rightarrow$ M5
				\\\hline  
\textbf{Master}	&0.403	&0.403	&0.079	&0.079	&0.289	&0.289\\
\textbf{Slave}	&0.513	&1.611	&0.247	&0.315	&0.352	&0.229\\
\textbf{DS}	&0.401	&0.402	&\textbf{0.152}	&0.117	&\textbf{0.280}	&0.297\\
\textbf{PDS}	&0.339	&\textbf{0.368}	&0.311	&0.241	&0.375	&0.366\\
\textbf{MFPI}	&\textbf{0.181}	&0.375	&0.262	&0.330	&0.684	&0.395\\
\textbf{MFPI-DS}	&0.400	&0.392	&0.183	&\textbf{0.113}	&0.300	&\textbf{0.280}\\
\textbf{MFPI-PDS}	&0.338	&0.480	&0.363	&0.281	&0.412	&0.411\\
				
			\end{tabular}
		}
	\end{center}
	
\end{table}

\end{document}